\documentclass[prl,superscriptaddress,showpacs,twocolumn]{revtex4}
\usepackage{srcltx}
\usepackage{amssymb,amsmath,amsfonts,graphicx}
\usepackage{hyperref}

\begin{document}
\title{Granular Solid Hydrodynamics:\\ Dense Flow,
Fluidization and Jamming}
\author{Stefan Mahle}
\affiliation{Theoretische Physik, Universit\"{a}t T\"{u}bingen,
Germany}
\author{Yimin Jiang}
\affiliation{Central South University, Changsha 410083,
China}
\author{Mario Liu}
\affiliation{Theoretische Physik, Universit\"{a}t T\"{u}bingen,
Germany}
\date{\today}
\begin{abstract}
Granular solid hydrodynamics, constructed to describe
quasi-elastic and plastic motion of granular solid, is
shown also capable of accounting for the rheology of
granular dense flow. This makes it a unified, though
still qualitative, hydrodynamic description, enabling
one to tackle fluidization and jamming, the hysteretic
transition between elasto-plastic motion and uniform
dense flow.

\end{abstract} 
\maketitle

A potentially catastrophic feature of granular media
is their variable capability to sustain external
stresses. As the mechanical stability of any structure
relies on this capability, it is important to have a
thorough understanding when and why it weakens, and
how it recovers. The transition from a solid-like
response to a liquid-like one is {\em fluidization};
and {\em jamming} denotes increasingly often the
reverse change. After a mud slide starts, after it
fluidizes, jamming is when it stops again -- at what
stress and density, and whether the village downhill
will be reached, are then questions of considerable
interests.

The two limiting states of the transition may be
referred to as {\em granular solid} and {\em uniform
dense flow}. In the first, the grains are deformed and
at rest, with all energy being elastic. In the second,
they jiggle, rattle, move macroscopic distances, and a
significant portion of the energy is kinetic. The
transition from the dominance of one energy to the
other may be gradual or abrupt, and has two possible
intermediate states: the uniform, ideally plastic {\em
critical state}~\cite{CSSM,PSM}, or {\em shear band},
the nonuniform path. A unified theory of these
limiting and intermediate states does not as yet
exist, though {\sc gsh} (for granular solid
hydrodynamics) seems close. It was originally
constructed to account for granular solid and its
elasto-plastic motion. Here, we demonstrate its
applicability to dense flow: {\sc gsh} displays broad
agreement with three existing theories on various
aspects of dense flow, and accounts for the data of
Savage and Sayed~\cite{Savage}.


The first of the three existing theories is by
Pouliquen et al.~\cite{pouliquen1}. Starting from the
insight that granular rheology in dense flows is
controlled by a dimensionless parameter
$\sim\dot\gamma/\sqrt P$ (where $\dot\gamma$ is the
shear rate, $P$ the pressure), they distilled two
locally applicable constitutive relations from
experiments and simulations, for the density $\rho$
and the friction angle $\sigma_s/P$ (with $\sigma_s$
the shear stress),
\begin{eqnarray}
\label{sb1} 1-\rho_{r}\sim \dot\gamma/\sqrt P, \quad
\frac{\sigma_s}{P}=\frac{\mu_1+\mu_2\,(\dot\gamma/\sqrt
P)^n}{1+(\dot\gamma/\sqrt P)^n},
\end{eqnarray}
where, with $\rho_{cp}$ the closed-packed density,
$\rho_r\equiv\rho/\rho_{cp}$ is the relative one.
$\mu_1,\mu_2$ denote, respectively, the friction angle
for $\dot\gamma\to0,\infty$. The authors took $n=1$,
though the difference to $n=2$ is subtle, as both
describe a gentle change from $\mu_1$ to $\mu_2$ with
$\dot\gamma$.

Earlier, Boquet et al.~\cite{Bocquet} developed a
continuum theory to account for their experiment.
Starting from the results of the kinetic theory for
inelastic hard spheres, they modified the density
dependence of the pressure $P$,  viscosity $\eta$ and
relaxation rate $\gamma$ to accommodate the higher
density in their system. They employ the Cauchy stress
$\sigma_{ij}$, and a balance equation for the granular
temperature $T_G$:
\begin{eqnarray}\label{sb2}
\sigma_{ij}=P-\eta v^0_{ij},\quad \partial
T_G/\partial t\sim \eta v_s^2-\gamma \, T_G,
\quad\text{with}
\\\label{sb3}
P\sim \frac{T_G}{1-\rho_r}, \quad \eta\sim
\frac{\sqrt{T_G}}{(1-\rho_r)^\beta},\quad \gamma\sim
\frac{\sqrt{T_G}}{(1-\rho_r)^\alpha},
\end{eqnarray}
where $\alpha=1$, $\beta$ between 1 and 2.5.
$v_{ij}\equiv\frac12(\nabla_iv_j+\nabla_jv_i)$ denotes
the strain rate, with $v^0_{ij}$ its traceless part,
and $v_s^2\equiv{v^0_{ij}v^0_{ij}}$ the scalarized
shear rate. (The notation $^0$ and $_s$ are used also
for other tensors below, such as the strain and
stress.) Solving both equations for a Couette cell,
the solution was found to agree well with their data.
This theory~II does not consider any elastic
contributions.

Considering shallow flows on an inclined plane and
rotating drums, Aranson and Tsimring identified the
hysteresis of transition, or the delay between jamming
and fluidization, as a key feature of granular
behavior~\cite{Aranson2}. Their theory~III treats the
Cauchy stress as the sum of two parts, a solid-like,
possibly elastic contribution
$\hat\varrho\sigma^s_{ij}$, and a rate-dependent fluid
one. A crucial variable is an order parameter
$\hat\varrho$ that is 1 for granular solid, and 0 for
dense flow. The authors take the friction angle
$\phi$, differently than above, as the ratio of the
solid stress components, and postulate a free energy
$f(\hat\varrho)$ such that granular solid,
$\hat\varrho=1$, is unstable for large shear stresses,
$\phi>\phi_1$; while dense flow, $\hat\varrho=0$, is
unstable for small ones, $\phi<\phi_0$. Both are
stable in the intermediate region,
$\phi_1>\phi>\phi_0$. This theory does not consider
variations in the density $\rho$, or in $T_G$, and
takes $\sigma^s_{ij}$ as an input from some other
theory. But its success provides a pivotal insight:
The viability, even appropriateness, of using a
partially bistable energy to account for the
hysteresis.

{\sc gsh} starts from the basic fact that grains with
enduring contacts are elastically deformed. Its
essential idea is that this deformation is slowly lost
when grains jiggle, as they briefly loose contact with
one another. Granular solid's complex elasto-plastic
behavior was shown to be a result of this simple
physics, assuming the dominance of the elastic energy.
Kinetic energy, or granular heat, is what underlies
the behavior of granular gas. So it seems obvious that
the behavior of dense flow results when both energies
are comparable, when the contribution to the stress
from granular temperature becomes equally important as
that from deformation.

{\sc gsh} was first employed to calculate static
stress distribution for various geometries, including
sand piles, silos, and point load, achieving results
in agreement with observation~\cite{ge1}. It was then
employed to consider slowly strained granular solid,
and found to yield response envelopes similar to those
from modern hypoplastic theory~\cite{granL3}.
Recently, the critical state -- generally considered a
hallmark of granular behavior -- was identified as a
steady-state, elastic solution of {\sc
gsh}~\cite{critState}: Although given as a simple
analytic expression, the solution realistically
renders the critical state and the approach to it,
including dilatancy and contractancy. Finally, the
velocity of elastic waves were calculated as a
function of the stress~\cite{elaWave}, and found to
agree well with experiments~\cite{Jia2009}.

{\sc gsh} consists of \textbullet~conservation laws
for the energy $w$, mass $\rho$, and momentum $\rho
v_i$, \textbullet~an evolution equation for the
elastic strain $u_{ij}$, and \textbullet~balance
equations for two entropy densities,  $s$ and $s_g$.
Two entropies are necessary, because granular media
display a {\em two-stage irreversibility}: Macroscopic
energy, kinetic and elastic, dissipates into
mesoscopic, inter-granular degrees of freedom, mainly
granular jiggling and the collision-induced,
fluctuating elastic deformation. After a
characteristic time, the energy degrades further into
microscopic, inner-granular degrees of freedom,
especially phonons. The granular and the true entropy,
$s_g, s$, account respectively for the energy of the
meso- and microscopic degrees of freedom. The elastic
strain $u_{ij}$ is the portion of the total strain
$\varepsilon_{ij}$ that deforms the grains and leads
to reversible storage of elastic energy. The
rest-frame energy density $w_0$ is a function of
$s_g,s,\rho,u_{ij}$ (though we shall neglect $s$, as
we are not interested in effects such as thermal
expansion at present). The conjugate variables are:
Granular temperature $T_g\equiv{\partial
w_0}/{\partial s_g}$, chemical potential
$\mu\equiv{\partial w_0}/{\partial \rho}$, elastic
stress, $\pi_{ij}\equiv -{\partial w_0}/{\partial
u_{ij}}$, and the gaseous pressure
$P_T\equiv\rho^2\left.{\partial(w_0/\rho)}
/{\partial\rho}\right|_{s_g/\rho,\dots}$. The elastic
stress $\pi_{ij}$ derives from granular deformation,
while $P_T$ is generated by granular temperature --
similar to the temperature generated pressure in a
gas. All conjugate variables: $(T_g, \mu,
\pi_{ij},P_T)$ are given once $w_0$ is.

In {\sc gsh}, the Cauchy stress $\sigma_{ij}$ [given
by momentum conservation, $\partial(\rho v_i)/\partial
t+\nabla_j(\sigma_{ij}+\rho v_iv_j)=0$] and the
balance equation for $s_g$ are given as
\begin{eqnarray}\label{sb4}
\sigma_{ij}=(1-\alpha)\pi_{ij}+P_T\delta_{ij} -\eta_g
v^0_{ij},
\\\label{sb5} \partial s_g/\partial t=
(\eta_gv_s^2 -\gamma T_g^2)/T_g.
\end{eqnarray}
Although a result of general principles, the
expression for $\sigma_{ij}$ is, remarkably, a simple
sum of the elastic stress, the gaseous pressure, and
the viscous stress, with $\eta_g$ the shear viscosity.
(Compressional flow is usually negligible. If not, one
needs to include the bulk viscosity.) For
elasto-plastic motion, only $(1-\alpha)\pi_{ij}$ is
important; granular gas is well accounted for by
$P_T\delta_{ij} -\eta_g v^0_{ij}$; dense flow needs
all three terms. $\alpha\approx0.8$ is a softening
coefficient that remains constant for all shear rates
considered in the present context. (It becomes smaller
only for ultra low shear rates, in ratcheting or
elastic waves). Note the similarity of Eq~(\ref{sb4})
to the above cited theories, with the difference that
theory~II ignores $\pi_{ij}$, and theory~III takes it
as given. In Eq~(\ref{sb5}), $\gamma$ is the
relaxation rate of $s_g$, accounting for the inelastic
collisions that occur when grains jiggle. The positive
term $\eta_g v_s^2\equiv\eta_g v^0_{ij}v^0_{ij}$
describes how grains, being sheared past one another,
start to jiggle in the process, leading to an increase
of $s_g$. From a more general point of view, this term
describes how the kinetic energy dissipates into
granular heat. In the stationary limit, for $\partial
s_g/\partial t=0$, we have
$T_g=v_s\sqrt{\eta_g/\gamma}$. (Only a uniform $T_g$
is considered -- more terms exist otherwise.)
See~\cite{granR2} for derivation and detailed
explanation.

Eqs~(\ref{sb4},\ref{sb5}) hold for the given set of
variables, independent of the granular material, or
the specific form of $w_0$. Material-specific
properties are encoded in $w_0(\rho,u_{ij},s_g)$, also
the transport coefficients: $\eta_g,\gamma$. We obtain
them from qualitative consideration, also comparison
to experiments and existing theories. (More
puristically,  one would of course like to obtain them
from simulation or microscopic calculations.) For dry
sand and glass beads, a simple energy expression, the
sum of the elastic energy $w_1(u_{ij},\rho)$ and
granular heat $w_2(s_g,\rho)$: $w_0=w_1+w_2$, has
turned out to be quite adequate as a first
approximation. Then $\pi_{ij}=-{\partial
w_1}/{\partial u_{ij}}$,
$P_T\approx\rho^2{\partial(w_2/\rho)} /{\partial\rho}$
(if one assumes $u_{ij}\ll1$, see~\cite{granR2}),
which is why stress and energy contributions are
simply linked: If $w_1$ dominates, only $\pi_{ij}$ is
important; while $P_T$ hinges on a sufficiently large
$w_2$. We have $w_1={\cal B}(\rho) \sqrt\Delta\,
[\Delta^2+u_s^2/\xi]$, with $\Delta\equiv
-u_{\ell\ell}$, $u_s^2=u^0_{ij}u^0_{ij}$. For a
granular system at rest, $w_1$ is the only energy, and
the elastic stress is the total stress. The mentioned
calculation of static stress distributions was carried
out using $w_1$~\cite{granR1}. Granular heat $w_2$ is
the lowest order expansion in $s_g$,
\begin{equation}\label{sb6}
w_2=\frac{s_g^2}{2\rho b},\quad T_g=\frac{s_g}{\rho
b}, \quad P_T=-\frac{T_g^2\rho^2}2\frac{\partial
b(\rho)}{\partial\rho}.
\end{equation}
The linear term vanishes because granular jiggling
dissipates and decreases toward zero, implying
$w_2(s_g)$ is minimal for $s_g=0$. Expanding also the
transport coefficients,
\begin{equation}\label{sb7}
\eta_g=\eta_0+\eta_1 T_g,\,\, \gamma=\gamma_0+\gamma_1
T_g, \,\, T_g=v_s\sqrt{\eta_1/\gamma_1},
\end{equation}
we take $\eta_0\ll\eta_1 T_0$, $\gamma_0\ll\gamma_1
T_g$, as is appropriate for any $T_g$ typical of
elasto-plastic motion and dense flow,
see~\cite{granR2}. Then the last of Eqs~(\ref{sb7})
holds, for $\partial s_g/\partial t=0$.

In theory~II, $T_{\rm G}$ is the energy per degree of
freedom, $w_2\sim T_{\rm G}$, while $w_2\sim s_g^2\sim
T_g^2$. We therefore identify $T_g\sim\sqrt{T_G}$, and
note the perfect agreement between
Eqs~(\ref{sb2},\ref{sb3}) and
Eqs~(\ref{sb4},\ref{sb5},\ref{sb6},\ref{sb7}). This is
important, because the $T_g$-dependence (in contrast
to the density dependence discussed below) is rather
fixed -- it is the result of the kinetic theory on one
hand, and the general consideration rendered above on
the other. (Note taking $w_2\sim T_g$ would disregard
the fact that $T_g$ dissipates, and $w_2$ is minimal
for $s_g,T_g=0$ in an adiabatic system. It is more
appropriate for ideal gas than the granular one.)

To consider the density dependence, we focus on
$1-\rho_r$ $\equiv1-\rho/\rho_{cp}$, which represents
a stronger dependency than $\rho$ if the sand is
dense, $\rho_r\approx1$. We take
\begin{equation}\label{sb8}
P_T=\frac{ab\rho_{cp}\rho_r^2\,T_g^2}{2(1-\rho_r)},\,\,
\eta_1=\frac{h_1\rho_{cp}}{({1-\rho_r})^{\beta}},\,
\gamma_1=\frac{g_1\rho_{cp}}{({1-\rho_r})^\alpha},
\end{equation}
with $\alpha=\frac12$, $\beta=\frac32$ to fit the
experimental results of~\cite{Savage} with respect to
the polystyrene beads. The expression for $P_T$
derives from $b=b_0(1-\rho_r)^a$, cf Eq~(\ref{sb6}).
Taking $b\sim\ln(1-\rho_r)$ would yield $P_T\sim
1/(1-\rho_r)$ exactly, as in Eqs~(\ref{sb3}), but also
leads to a divergent $s_g$, see~\cite{granR2}.
Assuming $a\approx0.1$ approximates the result, yet
avoids the problem. The coefficients $b_0,h_1,g_1$ are
(material dependent) numbers. Combining
Eq~(\ref{sb4},\ref{sb6},\ref{sb7},\ref{sb8}) and
denoting the shear rate as
$\dot\gamma\equiv\nabla_xv_y$ (hence
$\dot\gamma=\sqrt2\,v_s$ in simple-shear geometry), we
arrive at the final expressions for the pressure
$P\equiv\sigma_{\ell\ell}/3$ and the shear stress
$\sigma_s\equiv
({\sigma^0_{ij}\sigma^0_{ij}})^{1/2}$,
\begin{eqnarray}\label{sb9}
P=P_c+C_1\frac{\dot\gamma^2}{(1-\rho_r)^2},\quad%
\sigma_s=\Pi_c +C\frac{\dot\gamma^2}{(1-\rho_r)^2},
\end{eqnarray}
where $C_1=\frac14ab\rho_{cp}\rho_r^2\,{h_1}/{g_1}$,
$C=\frac12{\rho_{cp}}\sqrt{{h_1^3}/{2g_1}}$.
$P_c,\Pi_c$ denote the rate-independent, elastic
contributions, with $\Pi_c/P_c$ independent of
$1-\rho_r$, see the explanation below.

The first of Eq~(\ref{sb9}) may be written as
$1-\rho_{r}\sim
\dot\gamma/\sqrt{P-P_c}=\dot\gamma/\sqrt{P_T}$. It is
the same as Eq~(\ref{sb1}) if $P_c$ is neglected -- an
understandable mismatch, because the consideration of
theory~I involves inertia and confining pressure, but
neglects elasticity. The friction angle $\sigma_s/P$
is given by $\Pi_c/P_c$ for $\dot\gamma\to0$, and
$C/C_1$ for $\dot\gamma\to\infty$, which we may
respectively identify as $\mu_1,\mu_2$. The number $n$
of Eq~(\ref{sb1}) is 2 in {\sc gsh}. As mentioned, the
difference to $n=1$ is subtle for the friction angle
-- but less so if one look at the pressure or shear
stress individually. That both grow with
$\dot\gamma^2$ is in fact a behavior that already
Bagnold observed~\cite{Bagnold}. Finally, a note on
volume versus pressure control: Yielding $P, \sigma_s$
for given $\dot\gamma, 1-\rho_r$, Eqs~(\ref{sb9}) are
directly appropriate for experiments performed under
constant volume. If $P$ is fixed, one uses the first
calculate $\rho$, and rewrite the second as
$\sigma_s-\Pi_c=(P-P_c)C/C_1$, with a coefficient
$C/C_1$ that does not depend on $1-\rho_r$ -- though
still on $\rho$, a weaker function of $P$ and
$\dot\gamma$.

\begin{figure}[b]
\includegraphics[scale=.8]{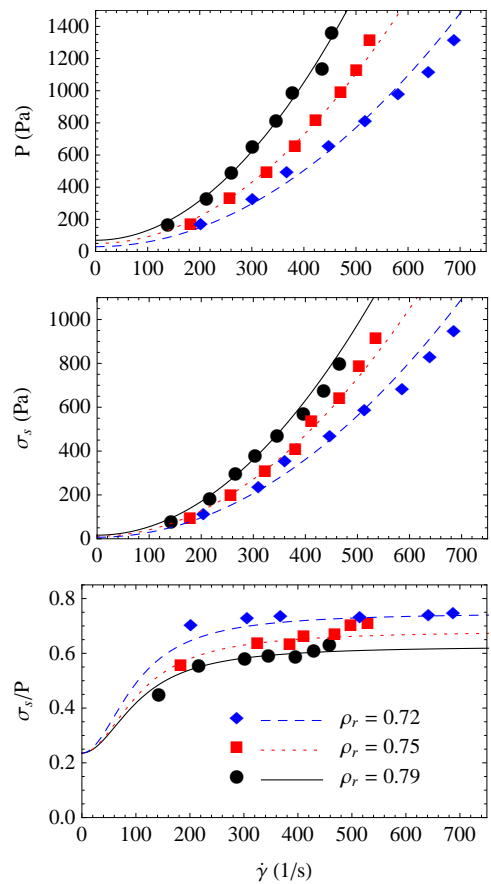}
\caption{Comparison of {\sc gsh} to the polystyrene
data of Savage and Sayed~\cite{Savage}, with Pressure
$P$, shear stress $\sigma_s$, and $\sigma_s/P$ given
as functions of the shear rate $\dot\gamma$. The first
two figures show the $\dot\gamma^2$-dependence, and
the third the convergence onto the weakly
density-dependent, high-rate limit $\mu_2=C/C_1$.
Diamonds, squares and circles are the experimental
points at the specified densities $\rho$. (Data for
the largest  $\rho$ are not used, because the authors
believe they may be plagued by ``finite-particle-size
effects.") The curves render Eqs~(\ref{sb9}), with
$h_1=3.1\cdot10^{-4}\sqrt{ab_0}$,
$g_1=121.7\sqrt{ab_0}^{3}$, $a=0.1$,
$\rho_{cp}=0.64\rho_{bulk}$.}\label{fig1}
\end{figure}
Returning to the exponent of $\gamma,\eta_1$, we note
$\alpha+\beta=2$ if $(\sigma_s-\Pi_c)/(P-P_c)$ is to
be independent of $1-\rho_r$; and $\beta-\alpha=1$, if
$P_T\sim(1-\rho_r)^{-2}$. Together, they imply
$\alpha=\frac12, \beta=\frac32$, as given above, see
Fig~\ref{fig1}. However, for glass beads of the same
experiment~\cite{Savage},
$P_T\sim\dot\gamma^2/(1-\rho_r)$, or $\beta=\alpha=1$
is more appropriate. In addition, the friction angle
decreases for increasing $\dot\gamma$ here, implying
$\Pi_c/P_c>C/C_1$, without contradicting any general
principle.

Next we discuss the elastic contributions:
$P_c,\Pi_c$. Applying a constant shear rate $v_s$ to
an elastic body, the shear stress will monotonically
increase -- until the point of breakage. Sand is
different and can maintain a constant stress,
$\sigma^c_{ij}$. This is the famous critical
state~\cite{CSSM,PSM} that has, for given density, a
unique, rate-independent stress value. Employing {\sc
gsh}, this is easy to understand: Because elastic
deformation $u_{ij}$ is slowly lost if the grains
jiggle, and because grains indeed jiggle when forced
to shear past one another, a shear rate $v_s$ not only
increases $u_{ij}$, as in any elastic medium, but also
decreases it. The critical state is the steady state
in which both processes balance each other, such that
the elastic deformation remains constant over time, in
spite of a finite $v_s$. As shown in~\cite{critState},
the stationary solution $u^c_{ij}$ depends on the
density, but not on $v_s$. The associated elastic
stress $\pi^c_{ij}\equiv\pi_{ij}(u^c_{k\ell},\rho)
=\pi^c_{\ell\ell}\,\delta_{ij}/3
-\pi^c_s\,v^0_{ij}/v_s$, characterized by two scalars,
$\pi^c_{\ell\ell}$ and $\pi^c_s$, is also independent
of $v_s$. Its contributions in Eq~(\ref{sb9}) are:
$P_c=\frac13(1-\alpha)\pi^c_{\ell\ell}$,
$\Pi_c=(1-\alpha)\pi^c_s$. Although both $P_c,\Pi_c$
depend on $1-\rho_r$, the ratio
$\Pi_c/P_c=\pi^c_s/\pi^c_{\ell\ell}$ does not.

We did not find any independent data on the critical
state of polystyrene beads, though that
from~\cite{Savage} indicate $P_c\approx50$~Pa,
$\sigma_c/P_c\approx0.25$, implying that the softer
polystyrene beads have a ${\cal B}\approx10^5$ Pa,
while the other coefficients retain their orders of
magnitude as given in~\cite{critState}. (Note:
$\pi_{ij}\sim{\cal B}$, and ${\cal
B}\approx5\times10^9$ Pa for sand.)

At lower shear rates, say for $v_s\lesssim$ 10
s$^{-1}$, the rate-dependent terms of $\sigma_{ij}$
are quadratically small, $P_T\sim T_g^2\sim v_s^2$,
$\eta_1T_g v^0_{ij}\sim v_s^2$, and may be neglected.
This is the reason the total stress is given by the
rate-independent critical state,
$\sigma_{ij}=(1-\alpha)\pi^c_{ij}$, for a fairly broad
range of shear rates, and why soil mechanic textbooks
emphasize the rate-independence of granular behavior.

We note that fluidization, as considered above, is
uniform and continuous, without anything resembling
``failure" or ``yield." Starting from a state of
isotropic stress, a sheared granular system will
approach the critical state, in the continuous way as
calculated in~\cite{critState}. The end state is, more
generally, given by Eqs~(\ref{sb9}), though the
difference to the critical state is evident only at
higher shear rates. There is an alternative path that
goes through an energetic instability, eg. the Coulomb
yield contained in $w_1(u_{ij})$, see~\cite{granR2},
which sets in when the ratio $\pi_s/ \pi_{\ell\ell}$
becomes too large. This transition is discontinuous,
non-uniform, and shear bands necessarily appear. We
shall consider it in a forthcoming paper.

Jamming, the reverse transition -- a drop of the shear
rate $v_s$ from a finite value to zero at given stress
-- is necessarily discontinuous. In contrast to the
authors of theory~III, however, we do not believe this
instability is marked by a lower bound of $\pi_s/
\pi_{\ell\ell}$, as elastic solutions are perfectly
stable at isotropic stresses, $\pi_s=0$. Rather,
jamming seems an instability that sets in when the
density is too high to enable a shear flow $v_s$.
Although $v_s$ is not a state variable, $T_g$ is, and
we have $T_g\sim v_s$ [cf. Eq~(\ref{sb7})] for any
processes slow enough for Eq~(\ref{sb5}) to have
reached its stationary limit. The appropriate
instability must therefore be in
$w_2(s_g,\rho)={s_g^2}/{2\rho b}$. If we substitute
$\hat b$ for $b$, we have $\hat P_T$ instead of $P_T$,
\begin{equation}\label{sb10}
\frac{\hat b}{b}=
\left[1+\frac{b_1}{1-\rho_r}\right],\quad \frac{\hat
P_T}{P_T}=1-\frac{(1-a)\,b_1}{a(1-\rho_r)}.
\end{equation}
With $b_1$ small, we may neglect the correction term
as long as $\rho_r$ is away from 1, and all results
above remain valid. For $\rho$ equal to
\begin{equation}\label{sb11}
\rho_{jam}=\rho_{cp}(1-2b_1/a),
\end{equation}
however, the convexity of $w_2$ with respect to $\rho$
is lost, and no finite value of $T_g\sim v_s$ is
stable. $\rho_{jam}$ is obtained from the condition:
$\partial^2 w_2/\partial\rho^2|_{s_g}=0$, or
equivalently, from $\partial^2
f_2/\partial\rho^2|_{T_g}\sim
\frac\partial{\partial\rho}(\rho^2\frac\partial
{\partial\rho}b)|_{T_g}=0$, where $f_2\equiv
w_2-T_gs_g=-\rho bT_g^2/2$. Eq~(\ref{sb11}) is the
result to lowest order in $b_1$ and $a$. Note
$\rho<\rho_{jam}$ imples a lower bound for $v_s$ if
the pressure is given instead of the density, as a
smaller $v_s$ will imply a larger $\rho$, see the
first of Eq~(\ref{sb9}). On a plane inclined by the
angle $\varphi$, the friction angle is
$\tan\varphi=\sigma_s/P$, with the angle of repose
given by $\varphi_{r}=\varphi(\rho_{jam})$. Since the
two terms $\sim\dot\gamma^2$ are negligible by then,
the angle of repose is given by the critical angle at
$\rho_{jam}$:
$\tan\varphi_{r}=\Pi_c(\rho_{jam})/P_c(\rho_{jam})$.
This is consistent with observation, because the
critical angle is necessarily smaller than the angle
at which Coulomb yield sets in. All these statements
are independent of the specific form of $\hat b$,
which may possibly prove inappropriate -- though the
case for an instability in $b(\rho)$ seems watertight.

Summary: Because {\sc gsh} is capable of accounting
for elasto-plastic motion, including the critical
state, and also for dense flow, fluidization and
jamming, we believe that this hydrodynamic theory,
conventionally based on conservation laws and
thermodynamics, is a viable candidate for a unified
theory of granular media.


\end{document}